\documentclass[a4paper]{article}
\usepackage{algorithm}
\usepackage{siunitx}
\usepackage{algorithmic}
\usepackage{microtype}
\renewcommand{\texttt}[1]{%
  \begingroup
  \ttfamily
  \begingroup\lccode`~=`/\lowercase{\endgroup\def~}{/\discretionary{}{}{}}%
  \begingroup\lccode`~=`[\lowercase{\endgroup\def~}{[\discretionary{}{}{}}%
  \begingroup\lccode`~=`.\lowercase{\endgroup\def~}{.\discretionary{}{}{}}%
  \begingroup\lccode`~=`-\lowercase{\endgroup\def~}{-\discretionary{}{}{}}%
  \catcode`/=\active\catcode`[=\active\catcode`.=\active\catcode`-=\active
  \scantokens{#1\noexpand}%
  \endgroup
}
\usepackage{textcomp}
\usepackage{balance}

\usepackage{itgspeech2021}    %% Include ITGSpeech2021 style
\usepackage{times}            %% Choose Times Roman font
\usepackage[english]{babel}   %% This is an English paper
\usepackage[utf8]{inputenc}%% use 'utf8' or 'ansinew' to support special characters (e.g., umlauts), depends on your editor!
\usepackage[T1]{fontenc}      %% use PS Type1 fonts
\usepackage[sort&compress,numbers]{natbib}	%% bibliography
\usepackage{amsmath,amssymb}
\usepackage{graphicx}
\usepackage[colorlinks=false,pdfborder={0 0 0}]{hyperref}
\usepackage{booktabs}
\usepackage{color}
\usepackage{units}

\usepackage{todonotes}      % this will show todonotes in text
\title{Federated Learning in ASR: Not as Easy as You Think}

\author{Wentao Yu \thanks{The authors contributed equally.}, Jan Freiwald \footnotemark[1], S\"oren Tewes, Fabien Huennemeyer, Dorothea Kolossa}

\address{Institute of Communication Acoustics, Ruhr University Bochum, Germany\\
Email: \texttt{\{wentao.yu, jan.freiwald, soeren.tewes, fabien.huennemeyer, \\ dorothea.kolossa\}@rub.de}}

\begin{document}

\maketitle
%The total length of the abstract is limited to 150 words. The abstract included in your paper and the one you enter during web-based submission must be identical. Avoid non-ASCII characters or symbols as they may not display correctly in the abstract book. 

\begin{abstract} % 150/150 words 
With the growing availability of smart devices and cloud services, personal speech assistance systems are increasingly used on a daily basis.
Most devices redirect the voice recordings to a central server, which uses them for upgrading the recognizer model.
This leads to major privacy concerns, since private data could be misused by the server or third parties.
Federated learning is a decentralized optimization strategy that has been proposed to address such concerns.
Utilizing this approach, private data is used for on-device training.
Afterwards, updated model parameters are sent to the server to improve the global model, which is redistributed to the clients.
In this work, we implement federated learning for speech recognition in a hybrid and an end-to-end model.
We discuss the outcomes of these systems, which both show great similarities and only small improvements, pointing to a need for a deeper understanding of federated learning for speech recognition.
\end{abstract}


\section{Introduction}
%Large-vocabulary automatic speech recognition (ASR) is used in different areas, such as in digital assistants like Amazon\textquotesingle s Alexa, or in healthcare \cite{durling2008speech}. %In recent time, the rapid development of computational devices allows large proportions of the populations to use a speech recognizer.
%Since speaker characteristics vary widely, data from many users is needed to optimize accuracy, and models are currently being trained on many thousand or even millons of hours of speech \todo[inline]{Reference needed}. 
In order to obtain training data that is optimally matched to the typical use cases and application environments of automatic speech recognition (ASR) systems, data of the users is sent to a central server.
In many commercial applications, like Amazon's Alexa, user speech recordings are decoded on a central server, where the service provider can incorporate the speech samples into their training and testing datasets.
This practice leads to massive datasets, spanning all kinds of different user characteristics and environments, which can be utilized to train highly optimized recognizer models.

In terms of privacy, this process also enables companies---or anyone, who may be able to access their data collections---to extract highly sensitive information about their users, such as identity, language, gender, personal conversation content and even emotional and health status. For example, in~\cite{galgali2015speaker}, the speaker\textquotesingle s age,
height and weight information are extracted from the audio signal. 

\begin{figure}[htbp!]
  \centering
  \includegraphics[scale=0.25]{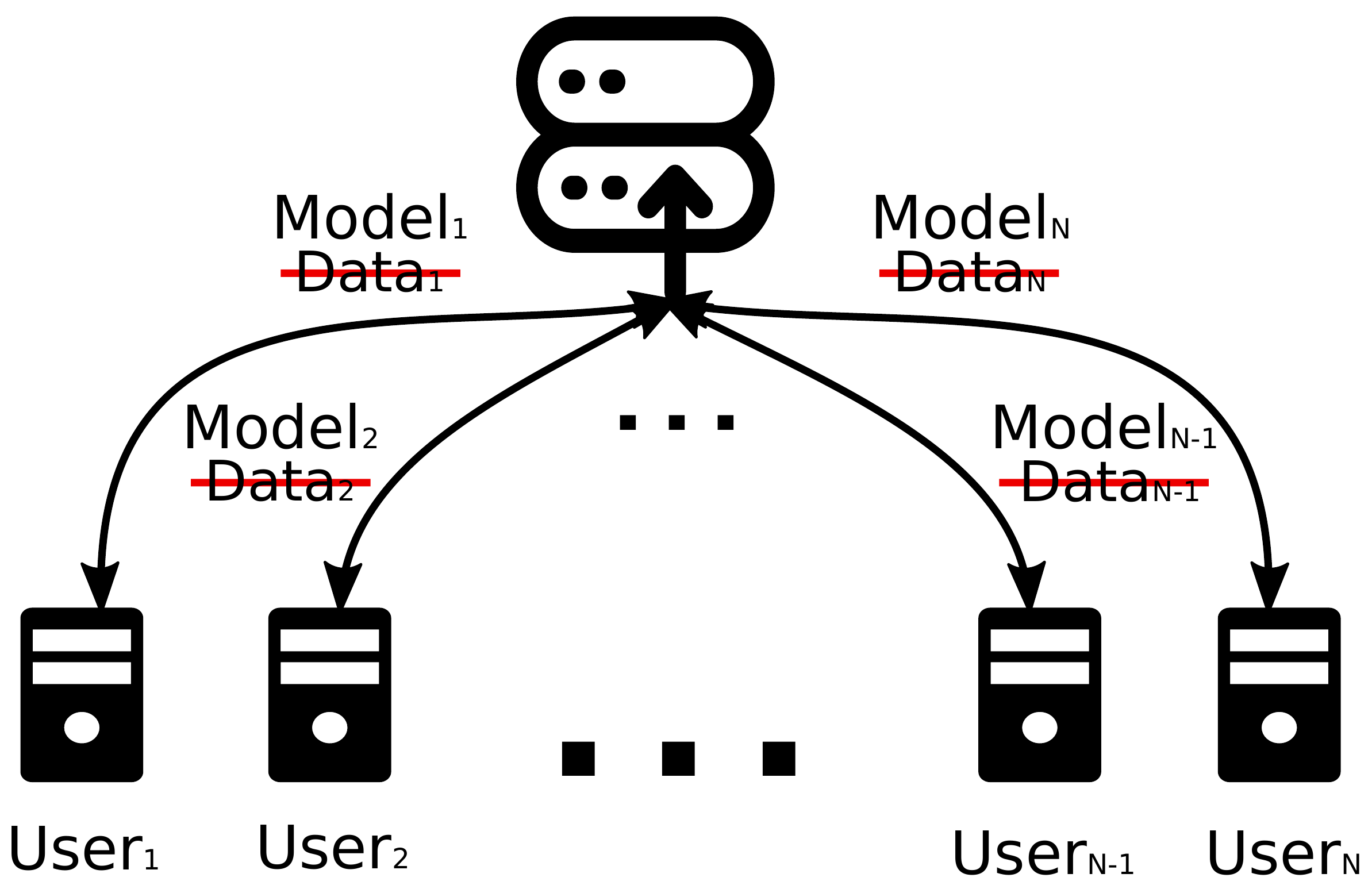}
  \caption{In a federated learning setup, no user data is transmitted to the central server. Instead, the server rolls out a model to the users, which is then updated locally and sent back to the server. The server calculates an updated model and repeats the cycle.}
  \label{fig:FL}
\end{figure} 

This leads directly into a dilemma: while big-data-driven AI shows exciting results in many different areas, centralized data processing leads to concerns about the security and the privacy of such systems and is not in line with the privacy-by-design principle stipulated in the European Union General Data Protection Regulation \cite{GDPR}. To counter this dilemma, the technique of federated learning (FL), as proposed by H. Brendan McMahan, et al.~\cite{mcmahan2017communication, konevcny2016federatedA, konevcny2016federatedB, mcmahan2021advances}, can be used.
%\todo[inline]{Names \cite{?}}
%(see e.g.~\cite{mcmahan2017communication, konevcny2016federatedA, konevcny2016federatedB,  hal-02406503v2}). The idea of the federated learning was proposed \todo[inline]{write by Names, [citation] instead of "`by Google"'}. 
The goal is to learn machine-learning models on data gathered on multiple private devices, while at the same time keeping all this re-training data private. For this purpose, federated learning proposes a decentralized optimization process. As shown in Figure~\ref{fig:FL}, in this way, it ensures user privacy by training on any private data locally, on-device, instead of sending private data to central servers. Each of the clients that are involved in re-training then sends a retrained model to the central server, and the central server updates its global model based on these user models, while all the clients' private data is kept on-device. In an optimal setup, the clients and the central server keep an appropriate communication frequency, maintaining a balance between communication cost and model performance~\cite{konevcny2016federated, hamer2020fedboost}.
%\todo[inline]{\cite{?}}.

In recent years, federated learning was applied successfully in many different tasks. In \cite{leroy2019federated, hard2020training}, federated learning is used in wake-word detection for a digital assistant. Federated learning was also applied in speech emotion detection~\cite{latif2020emotion}. In~\cite{hard2018federated}, federated learning is used for mobile keyboard prediction. Another example is in the medical area, where more and more data is available from hospitals, clinics, or patients, but strict privacy requirements limit its use. In this situation, federated learning promises a compromise that maintains privacy while allowing to improve the quality of healthcare through big-data-driven AI \cite{xu2021federated}.

%And even in medical areas, federated learning can allow for learning, while maintaining the patients\textquotesingle \ privacy~. %\todo[inline]{Please be more specific, what do you mean by "`And even in medical areas,"'?}
There are a few prior works on federated learning for large-vocabulary ASR. For example, in~\cite{gao2021end}, federated learning is applied with an end-to-end (E2E) model on the French set of the Common Voice dataset~\cite{ardila2019common}. In~\cite{guliani2020training}, an RNN-T architecture ASR~\cite{he2019streaming} is used on the LibriSpeech corpus, and \cite{cui2021federated} applies FL to a hybrid ASR model. Additionally, in \cite{Dimitriadis2020AFA} Dimitriadis et.al.~introduce a federated learning setup for ASR based on an end-to-end system.

In this paper, we directly compare the federated learning performance of two widely used ASR model types, i.e.~hybrid ASR based on the Pytorch-Kaldi system \cite{pytorch-kaldi} and the E2E transformer/CTC model, trained via ESPnet \cite{watanabe2018espnet}. We use the Librispeech corpus \cite{librispeech} for all experiments, which contains $\unit[960]{h}$ of speech data from 2484 speakers. Specifically, we consider a use case, where every client node only possesses data of one speaker, although scenarios with multiple speakers per client are conceivable (e.g.~cross-silo federated learning in  \citep{mcmahan2021advances}). 

We compare the performance and the communication cost of a number of federated learning strategies with different client-server communication frequencies and different model update strategies, finding that despite their large differences in approach and implementation, hybrid and E2E models show similar behavior, which raises the important question of how to achieve effective, decentralized learning in sequence-labeling tasks in general and ASR, specifically. To allow for easy reproducibility, we are also releasing the script that generates this dataset, as well as our training and test recipes\footnote{\url{https://github.com/rub-ksv/Federated-learning-ASR}}.

The paper is organized as follows: The idea of federated learning and the different update strategies are detailed in Section~\ref{setup}, followed by a brief introduction to the PyTorch-Kaldi and ESPnet models. Section~\ref{sec:dataset} describes our reorganization of the Librispeech dataset, which allows us to evaluate FL strategies in a meaningful manner. We present the results in Section~\ref{results} and discuss the outcomes and the questions that they raise in Section~\ref{Conclusions}.

%First we give an overview of related works and software used for this investigation. Then we explain our system setup and the changes we made to the librispeech dataset.

%\input{parts/relate.tex}

\section{System overview}\label{setup}
In this section, we take a look at federated averaging in general. Our experiments rely on two ASR toolkits, PyTorch-Kaldi and ESPnet. Hence we give a short introduction to both frameworks and describe our training configurations for both. 

\subsection{Federated averaging}\label{sec:fedavg}
Federated averaging (\textit{FedAvg}) \cite{mcmahan2017communication} is a central algorithmic step in federated learning. As shown in Algorithm~\ref{fadavg}, a pretrained model $\mathbf{W}^0$ initializes all clients at the beginning of the learning process.
This is followed by an iterative optimization loop, which contains two steps per round: 
%the local optimization, which is performed by each client and the global optimization step, which is executed by the server. In each of the traininig rounds, selected client models are initialized by the previous global model. Then the client models are updated with each clients private dataset. Finally, the \textit{FedAvg} is performed to update the global model.
the first step is the local optimization, which is performed by each client.
Each client's model is initialized by the previous global model $\mathbf{W}^{i-1}$. To obtain the new client-side local model, each client updates their model with their own private dataset, which is kept on-device.
The second step is the \textit{FedAvg} step, which is applied over all selected client models. 
The averaged model is considered as the new global model in the next round.
In contrast to other works, averaging the accumulated gradient updates (e.g~\cite{mcmahan2017communication, guliani2020training}), we directly average the model parameters at the end of each round, which leads to the same outcome with an easier implementation:

\begin{equation}  \label{paramupdat}
\begin{aligned}
\mathbf{W}^{i} &=\sum_{n}^{}\alpha_n\cdot \mathbf{W}^{i}_{n}\\ 
%&=\alpha_1\cdot (\mathbf{W}^{i-1} +\Delta \mathbf{W}^{i}_{1}) + \cdots 
% +\alpha_N\cdot (\mathbf{W}^{i-1} +\Delta \mathbf{W}^{i}_{N}) \\ 
&=\sum_{n}^{}\alpha_n\cdot \mathbf{W}^{i-1} + \sum_{n}^{}\alpha_n\cdot \Delta \mathbf{W}^{i}_{n} \\
&=\mathbf{W}^{i-1} + \sum_{n}^{}\alpha_n\cdot \Delta \mathbf{W}^{i}_{n},
\end{aligned}
\end{equation}
%\todo[inline]{please replace all $\textit{\textbf{W}}$ by $\mathbf{W}$ in the entire document}
%\todo[inline]{need to define $\alpha_n$ and $\Delta \mathbf{W}^{i}_{n}$ }
where $\alpha_n$ are the \textit{FedAvg} weights described in Algorithm~\ref{fadavg} and
$\Delta \mathbf{W}^{i}_{n}$ is the model parameter update for client $n$ in the $i$-th iteration.

In Algorithm~\ref{fadavg}, the federated learning round can be performed with different communication frequencies. We have identified the following levels for defining these:
\begin{itemize}
\item \textbf{Batch-level (B)}: The communication is always performed after $K$ mini-batches. In every communication round, the selected client examines whether there are untrained utterances for this epoch. If the untrained utterances of the current speaker do not suffice for $K$ mini-batches, the batch size is reduced. The client stops sending a new model, when there is no more training data in the current epoch.
After an epoch is completed, a new round begins with the complete set of clients participating again. %\todo[inline]{why call this iteration-level, if definitions are made for batches? why not call it batch-level?}

\item \textbf{Epoch-level (E)}: the update is performed with the selected clients' models after every epoch.
%\todo[inline]{this sentence is not clear: "`We can select all clients' models for the \textit{FedAvg} update."' Can we also \emph{not} select all clients models?}
In the ESPnet setup, we can select all client models for the \textit{FedAvg} update, or alternatively, we can randomly select $N$ client models for updating. In PyTorch-Kaldi, we can also perform updates after a fixed fraction of an epoch.
% s instead of a fixed number of batches.[verwirrendlich, weil vorher gar nicht die Rede von Batches war]
%, because every epoch is divided into $N$ chunks. [das würde ich weglassen - auch wg. Namenskonflikt mit den N clients]

\item \textbf{Convergence-level (C)}: there is only one update, which is performed with the selected clients' final models, after convergence.
%The client-side local models are trained until they converge.
%So the final model is only updated once using \textit{FedAvg}.
\end{itemize}

The \textit{FedAvg} weights $\alpha_n$ in Algorithm~\ref{fadavg} can be defined as:
\begin{itemize}
\item \textbf{Mean (M)}: $\alpha_n = \frac{1}{N}$, where $N$ is the number of the selected clients.
\item \textbf{Weighted (W)}: $\alpha_n = \frac{|\textbf{D}_n|}{\sum_{i}^{}|\textbf{D}_i|}$, where $|\textbf{D}_n|$ is the number of utterances of the client-side dataset for client $n$.
\end{itemize}

\begin{algorithm}[htp] 
\caption{Federated averaging. There are $N$ clients. The initial dataset is $\textbf{D}_0$. The federated learning training-set $\textbf{T}^i_n$ for client $n$ in round $i$ is derived from the complete client-side training set $\textbf{D}_n$.}
\label{fadavg}
\begin{algorithmic}[1]
\STATE $\mathbf{W}^0$ = \textit{train\_initial\_model}($\textbf{D}_0$)
\FOR{round  $i \in [1, 2, 3, \cdots]$ }
\FOR{client $n \in [1..N]$}
\STATE $\mathbf{W}^{i}_{n}$ = $\mathbf{W}^{i-1}$
\STATE $\textbf{T}^i_n=\textit{subset}(\textbf{D}_n, i)$
\STATE $\mathbf{W}^{i}_{n}$ = \textit{train\_model}($\textbf{T}^i_n$, $\mathbf{W}^{i}_{n}$) 
\ENDFOR
\STATE $\mathbf{W}^{i} =\sum_{n}^{}\alpha_n\cdot \mathbf{W}^{i}_{n}\text{, where } \sum_{n}^{}\alpha_n = 1$
\ENDFOR
\end{algorithmic}
\end{algorithm}
%And the selected clients could be varied from randomly selected $N$ clients to contain all clients. 
\vspace*{-2mm}
\subsection{PyTorch-Kaldi}
PyTorch-Kaldi\cite{pytorch-kaldi} allows for high flexibility when using DNNs in training hybrid models for use with Kaldi. For this purpose, it provides a model configuration interface, as well as training recipes.

Our overall setup is divided into three parts, Kaldi training, PyTorch-Kaldi training, and \textit{FedAvg}.
First, the features of all relevant datasets are extracted, the corresponding GMM models are trained, and alignments are obtained in Kaldi \cite{povey_ASRU2011}, using MFCCs with first and second derivatives and cepstral mean and variance normalization on a per-speaker basis. 
In the second part, %the configuration files are created in order to pass all the audio features and the alignments, which are required for neural network training to PyTorch-Kaldi \cite{pytorch-kaldi}.  
we use PyTorch-Kaldi to train a hybrid DNN-WFST model with the PyTorch library. We use the standard MLP structure that is provided with many of the examples in PyTorch-Kaldi, a feed-forward network with a context size of 5 frames~\cite{pytorch-kaldi}. This results in the initial model $\mathbf{W}^0$, which is only learned on the \texttt{initial} dataset.
After the server-side training is thus completed, federated learning is performed as described in Algorithm~\ref{fadavg}. Here, client-side trainings only re-train the DNN of the hybrid model, leaving the WFST unchanged.
%Thereafter, the new model is redistributed to all the clients.
%To evaluate the model performance, the Word Error Rate (WER) is calculated.

%To get the federated learning training started, we performed a pretraining of an initial hybrid, based on the \texttt{initial} dataset.
%As common with Pytorch-Kaldi, we used a GMM-HMM triphone model to aligned our training datasets. [das sollte oben stehen] 
%This aligner model was trained on the \texttt{initial} training set and applied in a forced alignment to every subset. 
% We used MFCC+DELTA+DELTA-DELTA features with CVMN on a per speaker basis as features. 

%\todo[inline]{Dorothea: what do you mean by 'compare the setup...' why is that sentence necessary? Is this a standard network topology or is it simpler than usual? If it is a standard, which recipe is it modeled after? }

The initial model is trained for 24 epochs and distributed to all clients, followed by 12 federated learning epochs with different communication frequencies.
In order to use the standard training script for federated learning, we changed it to pause and save a checkpoint model after a specifiable number of data chunks; every clients' training data is divided into $n=8$ chunks.
This allows us to train our network on each of our clients, then pause the training, average every updated model and redistribute the result to all clients. In this way, we can perform (potentially partial) epoch-level training, as described in Section~\ref{sec:fedavg}.

%\todo[inline]{what do you mean by partial E-level training? is there a way to express this in the terminology from the previous subsection? if not, we may need to update terminology to reflect what we did (Wentao's unstand: maybe you mean I-level update, but you should calculate, how many mini-batches are trained, until 0.5 epoch and 0.25 epoch)}
The scripts are based on training recipes provided for Kaldi~\cite{povey_ASRU2011} and PyTorch-Kaldi~\cite{pytorch-kaldi}. We used, among others, the WSJ, Librispeeech and CommonVoice recipes as a guide for structuring and parameterizing the training and decoding of our models.

\subsection{ESPnet setup}
The ESPnet end-to-end model \cite{watanabe2018espnet} uses a sequence-to-sequence (S2S) transformer model with connectionist temporal classification (CTC) optimization. As described in~\cite{graves2006connectionist}, 
%\todo[inline]{this should cite an original ctc paper, not one from 2019}, 
CTC learns to align features and transcription optimally, which leads to fast convergence.
The model is updated using a linear combination of CTC and S2S loss terms as the objective function
\begin{equation} \label{jointtraining}
L=\eta  \cdot \textrm{log}\ p_\text{CTC}(\textbf{s} |\textbf{o}) +(1-\eta  )\textrm{log}\ p_\text{S2S}(\textbf{s} |\textbf{o}),
\end{equation}
with $\textbf{s}$ as the states and the constant hyper-parameter $\eta $, which controls the strength of both terms. The
joint S2S-CTC model performs token-by-token decoding, until the end-of-sentence symbol (EOS).
In each step, the log-poste\-riors of the current token from the Transformer and CTC models are linearly combined with an RNN-language model $p_\text{LM}(\textbf{s}_t)$ 
\begin{multline} \label{jointdecoding1}
\textrm{log}\ p^{\ast }(\textbf{s}_t |\textbf{o}) = \eta \ \textrm{log}\ p_\text{CTC}(\textbf{s}_t |\textbf{o}) +  (1-\eta )\ \textrm{log}\ p_\text{S2S}(\textbf{s}_t |\textbf{o}) + \\ \theta\ \textrm{log}\ p_\text{LM}(\textbf{s}_t) ,
\end{multline}
where $\theta$ controls the contribution of the language model and is set to 0.3. 

All models use 80 log-mel filterbank features together with pitch, delta pitch, and the probability of voicing. These 83-di\-men\-sio\-nal features are extracted using \SI{25}{\ms} frame size and \SI{10}{\ms} frame shift. %\todo[inline]{use units package to get right distance between number and unit}. 

%The learning rate and the optimizer settings are the same as in \cite{vaswani2017attention}. The learning rate schedule follows the function: 
%\begin{equation} \label{lr}
%\delta  = K \cdot 256^{-0.5}\cdot \textrm{min}%%(L^{-0.5}, L \cdot M^{-1.5}),
%\end{equation}
%where $K$ is a factor, $L$ is the number of the current steps, and $M$ represents the defined warm-up steps. In Equation~\ref{lr}, the learning rate first increase linearly during the warm-up steps,  then is decrease proportionally to the inverse square root of the step number. The Adam optimizer \cite{kingma2014adam} with $\beta _1$ = 0.9, $\beta _1$ = 0.98 and $\epsilon$ = $10^{-9}$ is used for all experiments.

For the language model, we use a 4-layer recurrent neural network, which predicts a character at a time and receives the previous character as the input. Each layer consists of 2048 units. The language model is trained on the complete LibriSpeech corpus for 20 epochs using SGD. 

%%%%%%%%

The dataset is divided into a server-side dataset and user-side datasets (see Section~\ref{sec:dataset}). The \texttt{initial} training set is used to train the initial model $\mathbf{W}^0$. All ESPnet experiments are trained using NVIDIAs Volta-based DGX-1 multi GPU system. The initial model is trained on 3 Tesla V100 GPUs, each with \SI{32}{GB} memory, for 100 epochs at a batch size of 32. The upper bound reference model, trained on the \texttt{complete} set, uses the same model configuration and hyperparameter settings. For the initial and the reference model, training takes two days, and four days, respectively. In the federated learning stage, the batch size is set to 8. The hyper-parameter $\eta$ is set to 0.3 during training and $\eta=0.5$ during decoding.%\todo[inline]{how is $\eta$ set?}

 %In our setup, the initial model is distributed to all clients, so the learning rate $K$ and $M$ for each client are much smaller than in initial training, 1.65  and 800, respectively. The batch size is 8.  Based on the ESPnet implementation, the batch size might vary depending on the input and output length of the utterances, meaning that for utterances with input length $\geq 512$ or output length $\geq 150$ the batch-size would automatically be reduced. 

For C-level averaging, where each client trains their model to convergence, early stopping is used, and triggered if the evaluation accuracy does not improve over 10 epochs. Both M-averaging and W-averaging are applied in the C-level experiments. 
As described in Section~\ref{sec:fedavg}, we either use all clients to update the model or randomly select $N$ of them in each round.
To asses the impact of this choice, we repeat the E-level training with $N = 100$, $N = 350$ and $N = all$. 
%All federated learning strategies are trained with a single GPU. 
%\todo[inline]{I deleted the next sentence: The client local models are trained sequentially.}
% The client local models are trained sequentially.
%\todo{What is the point of this? Is this relevant? ->} The clients and server communicate more often, the communication cost became heavier. The C-level training takes almost 4 days; E-level training time depends on how many clients are participants in every epoch, but it takes longer time than C-level training. For example, training with $N = all$ costs 8 hours for one epoch; I-level updating takes the longest time. It takes 2 days for one epoch.

\section{Dataset}\label{sec:dataset}

For this work, we reorganized the Librispeech\cite{librispeech} dataset, creating dedicated training, test and development sets to simulate a realistic federated learning environment.

Therefore, we created a server-side training set for our initial and alignment models (\texttt{initial}). Additionally, we created user datasets, which contain training, development, and test sets for every client. All 1372 clients are disjoint from the initial training set and from each other. 
%Therefore the \texttt{federated} train-set is grouped by speaker id. 
Furthermore, the union of the \texttt{federated} and \texttt{initial} training sets is termed \texttt{complete} and used to obtain an upper performance bound, corresponding to what would be seen if all data were located on a central server, i.e.~with standard, non-privacy-preserving learning. 

For testing, we created three sets: \texttt{unseen} is composed of previously unseen speakers, \texttt{initial} contains speakers from the training set of the initial model, and \texttt{federated} contains speakers from the federated training set, so it simulates clients that can profit from their data contribution.

Statistics of this dataset partitioning are shown in Tab.~\ref{tab:utts}.
%Our Pytorch-Kaldi and ESPnet setups use identical datasets. 

\begin{table}[hbp!]
\centering
\caption{Dataset composition of all training and test sets. The proportion of male and female speakers is close to 50\%.}
\begin{tabular}{c|rcc}
\toprule
&Dataset &Utterances &Duration [h] \\
\midrule
%\textbf{Train-sets} & & \\
&\texttt{initial} & 37768 & 134.5\\
Training &\texttt{federated} &  110570 & 367.8\\
&\texttt{complete} & 148338 & 502.3\\
%\texttt{fl/*/dev}   &  37031 & 123.4\\
%\texttt{fl/*/test}  &  37593 & 124.9\\
%\midrule
%\textbf{Dev-sets} & & \\
%\texttt{unseen}  &  490 & 1.2\\
%\texttt{initial}  &  487 & 1.7\\
%\texttt{federated}  &  682 & 2.3\\
\midrule
%\textbf{Test-sets} & & \\
&\texttt{initial}  &  487 & 1.7\\
Test &\texttt{federated}  &  682 & 2.3\\
&\texttt{unseen}  &  476 & 1.1\\
\bottomrule
\end{tabular}

\label{tab:utts}

\end{table}

\vspace*{-2mm}
\section{Results}\label{results}

\subsection{Hybrid ASR}
The results of the PyTorch-Kaldi experiments are summarized in Table~\ref{tab:pykaldi_res}. As expected, the performance improves along the training phases, from the GMM-HMM alignment model via the initial model to the FL-models. In all cases, the upper bound reference model performs best. FL-training was performed at the E-level, where the number in parentheses shows the number of epochs after which communication takes place. We find that more frequent update rounds of \textit{FedAvg} slightly improve the overall results, with best performance observed for updates after every $1/2$ or $1/4$ of an epoch. The highest improvement through FL-training is seen on the \texttt{federated} test set, which contains utterances of the FL speakers. The recognition quality of the other test sets does not degrade after FL-training.

\vspace{-5pt}
\begin{table}[ht]
\centering
\caption{WER (\%) of hybrid models trained conventionally and using E-level federated learning, with associated communication cost [GB]. The final two rows show the WERs of upper bound reference models trained on the \texttt{complete} set. Ref.~1 learns its decision tree on the \texttt{complete} set; Ref.~2 uses the decision tree of the initial model.}
\setlength\tabcolsep{2.0pt}
\vspace{2pt}
\begin{tabular}{r|cccc}
\toprule
% & \multicolumn{3}{c}{WER [\%] of decoding-set \texttt{final/...}} \\
 & \small\texttt{unseen} & \small\texttt{federated} & \small\texttt{initial} & cost[GB]\\
\midrule
Aligner    & 32.76 & 33.22 & 26.96 & - \\
Initial    & 19.06 & 18.18 & 14.05 &- \\
\midrule
E(2)-M     & 19.08 & 17.86 & 14.05 & 326\\
E(1)-M     & 19.01 & 17.79 & 14.05 & 651\\
E(1/2)-M   & 19.02 & \textbf{17.67} & \textbf{13.90} & 1302\\
E(1/4)-M   & \textbf{18.94} & 17.71 & 13.94 & 2604\\
\midrule
Ref.~1       & 17.77 & 15.78 & 13.24 & -\\
Ref.~2     & 17.70 & 16.13 & 13.29 & -\\
\bottomrule
\end{tabular}
\label{tab:pykaldi_res}
\vspace{-10pt}
\end{table}

\subsection{End-to-end learning}
Table~\ref{tab:esp_res} shows the experimental results using the ESPnet toolkit. Analogously to the PyTorch-Kaldi experiments, the upper-bound reference model provides best performance. For C-level FL, both averaging methods are applied, but as there is no large difference between W- and M-averaging, we applied only W-averaging in the later experiments.
%because the number of utterances of each client is distributed approximately evenly in our dataset.

The E-level experiments (E-100-W, E-350-W, and E-all-W) are trained on randomly selected $N = [100, 350, \text{all}]$ speakers per round. %Comparing the influence of the amount of the participated clients on FL training. 
As expected, we see that a larger number of clients $N$ tends towards a better performance.

We also tested the B-level strategy (\textbf{B}), updating the model after every minibatch, which we hoped would give optimal performance, as it best approximates the situation in non-FL training. However, while the communication cost became higher as expected, there are slight performance improvements at best, which do not justify this extra effort.

\begin{table}[htbp]
\centering
\caption{WER (\%) of end-to-end models trained on the \texttt{initial} set and subsequently applying FL. Last row: WER of a reference model trained on the \texttt{complete} set. \textrm{C}, \textrm{E}, \textrm{B} represent \textrm{C}-, \textrm{E}-, and \textrm{B}-level updates;  \textrm{M} and \textrm{W} indicate M- and W-averaging.}
\setlength\tabcolsep{2.0pt}

\begin{tabular}{r|cccc}
\toprule

 & \small\texttt{unseen} &\small \texttt{federated} & \small\texttt{initial} & cost [GB]\\
\midrule
Initial & 8.98  & 6.94  & 4.23  & - \\ %&6.72\\
\midrule
C-M &9.00 &6.88 &4.18 &317 \\
C-W &9.01 &6.88 &4.15 &317\\
E-100-W    & 8.91 & 6.89 & 4.17 & 925\\
E-350-W    & 8.87 & 6.91 & \textbf{4.14} & 1618\\
E-all-W    & \textbf{8.81} & 6.86 & 4.17 & 4758\\
B-W &8.93 &\textbf{6.81} &4.15 & 13319\\
\midrule
Ref & 6.08 & 3.37 & 3.07 & - \\ %&4.18\\
\bottomrule
\end{tabular}

\label{tab:esp_res}
\end{table}

%Table~\ref{tab:esp_res} also gives an overview of the communication costs for different strategies.
%In our experiments, every exchanged model is 121213 KB. For C-level model update, considering both up-and download cost, there is $= 317.2$ GB data exchange. For E-level update \textbf{E-100}, \textbf{E-350}, and \textbf{E-all} have 924.8, 1618.4, and 4758.0 GB data exchange, respectively. The communication cost in the I-level update is the heaviest, it has 13318.94 GB data exchange. %9 * (900 + 4 * 1372) * 2 * 121213 KB 

We also see that the C-level updates, most effective in terms of communication cost, can still improve the recognition accuracy compared to the pretrained model, but the improvement does not reach that of the E-level updates. E-level updates with $N$ randomly selected speakers may also be closest to a realistic scenario, as we would not wish to ask every client to participate in every one of the model updates.

%\begin{table}[htp]

%\setlength\tabcolsep{9.0pt}

%\begin{tabular}{r|ccc}
%\toprule

%\textbf{Setup} & epochs & frequency (times) & cost %(GB) \\
%\midrule
%C & - & 2744  & 317.20  \\
%E-100  & 40 & 8000  & 924.78  \\
%E-350  & 20 & 14000  & 1618.37  \\
%E-all  & 15 & 41160  & 4758.00  \\
%I  & 9 & 115218  & 13318.94  \\
%\bottomrule
%\end{tabular}
%\caption{Communication frequency and cost in both up-%and download for different strategies of the ESPnet. %Epochs indicate the number of the training epochs. %Each model is 121213 kB.}
%\label{tab:cost}
%\end{table}

%\begin{table}[htp]
%\label{tab:Espnetresults}
%\setlength\tabcolsep{3.5pt}
%\caption{This table shows the WER (\%) improvements of different ESPnet models before and after FL-training. }
%\begin{tabular}{r|ccc|c}
%\textbf{Setup} & \texttt{test} & \texttt{fl-test} %& \texttt{pre-test} &mean\\
%\midrule
%Initial model  & 0.00 & 0.00 & 0.00 &0.00\\
%\midrule
%C + M &-0.02 &0.06 &0.05 &0.03\\
%C + W &-0.03 &0.06 &0.08 &0.04\\
%E + W + $N=100$    & 0.07 & 0.05 & 0.06 &0.06\\
%E + W + $N=350$    & ? & ? & ? &?\\
%E + W + $N=all$    & 0.17 & 0.08 & 0.06 &0.11\\
%I &0.05 &0.13 &0.08 &0.09\\
%\midrule
%Upper bound model & 2.90 & 3.57 & 1.16 &2.54\\
%\bottomrule
%\end{tabular}
%\end{table}

\section{Conclusions}\label{Conclusions}

We have shown that federated learning can slightly improve the recognition rates in automatic speech recognition, both using hybrid and end-to-end models. %However, the improvements are not significant. 
We suspect that one reason for the disappointing performance is to be found at the heart of the federated averaging approach. During FL-training, federated averaging leads to smaller gradients and, hence, to slow improvements.
%{\color{blue}Nevertheless, we tested larger learning rates with a preliminary ESPnet-setup, but found no improvments.} [dorothea: das würde ich hier lieber weglassen - ich finde die Idee, die Lernraten größer zu wählen, nicht unbedingt so vielversprechend, und glaube eher, dass die Strategie des Clustered FL helfen könnte. Aber das bekommen wir hier mit dem wenigen Platz auch nicht mehr richtig erklärt.]
While it is possible that better performance may be attained by engaging more clients and allowing for more model updates, this comes at an excruciating communication cost. Therefore, we believe that it is necessary to focus on more effective collaborative learning strategies, e.g., based on suitable client selection criteria or on re-training only parts of the networks.

Furthermore, McMahan et al.~\cite{mcmahan2021advances} point to an inherent problem of the federated learning approach, when it is applied on non-Independent, Identically Distributed (non-IID) client data. Since on-device training data is naturally grouped by speaker, there are different statistical properties in every client, which may degrade the model performance. To counter this problem, a cross-silo training approach can be beneficial and will be subject of our future investigations. %{\color{blue} Another way to improve the performance is using  
Federated Variational Noise (FVN) will also be investigated, which has shown its value in recovering FL-ASR performance despite non-IID client data \cite{guliani2020training}.

In addition to this global view, it is also interesting to look at the model architectures in detail:

For hybrid models, federated learning imposes several difficulties and impracticalities. Due to the comparatively smaller size of the initial training data, decision tree learning results in a smaller output layer size of the FL-trained models in contrast to the full model. This also hints at a similar issue in the calculation of the language model, which we did not consider here, but which would also likely have a smaller size and larger perplexity if it is not updated---while simultaneously, any updates of the language model by nature lead to a leakage of private speech content.

We also found that fewer iteration steps between \linebreak \textit{FedAvg} improve the quality of the model, but impose a heavy computational burden. %In this particular case the constant startup-time of the python training script heavyly impacts the overall training time and makes it virtually impossible to calculate finer grained updates.
Finally, the alignment model is not updated at the moment, and no realignments of client training data were performed after model averaging, which may also be a factor in the observed performance issues. This additionally raises the interesting question, in how far---explicit or implicit---alignment is an issue in federated learning for time-series modeling in general. 

In contrast, the end-to-end-model suffers neither from an impacted decision-tree nor from a missing explicit re-alignment. We conducted many experiments to assess the influence of different parameters and hyper-parameters of federated training, finding that allowing more participants in each round of model averaging yields better performance, and that in trying to balance the performance and cost, epoch-level averaging appears most promising.

All in all, however, we see the outcomes of our experiment as a pointer to underlying issues of the idea of federated averaging, when it is applied to complex, highly variable time series data, as seen in large-vocabulary speech recognition. We are releasing our source code in the hope of setting a starting point for further investigations into the important, yet also challenging question of how to achieve a reasonable balance of performance and cost in privacy-preserving machine learning.

{\small\section*{Acknowledgements}
This work was funded by the PhD School ``SecHuman - Security for Humans in Cyberspace'' by the federal state of NRW, and by the German Federal Ministry of Education and Research (BMBF) within the ``Innovations for Tomorrow’s Production, Services, and Work'' Program (02L19C200), a project that is implemented by the Project Management Agency Karlsruhe (PTKA). The authors are responsible for the content of this publication.}
%This work is also financed by the German Research~Foundation (DFG) under under grant number KO3434/4-2.}

%\centering\includegraphics[width=\columnwidth]{logos_bmbf_dfg.png}

\bibliographystyle{IEEEtran}
\balance
\bibliography{mybib}

\end{document}